\documentclass{winnower}
\usepackage{color} 
\begin{document}
\setlength{\parindent}{2em}  
\setlength{\parskip}{0.2em}  
\renewcommand{\baselinestretch}{1.0}  

\title{\textbf{Numerical analysis of a mechanotransduction dynamical model reveals homoclinic bifurcations of extracellular matrix mediated oscillations of the mesenchymal stem cell fate}}

\vspace{4in}
  
\author{Katiana Kontolati}
\affil{School of Applied Mathematical and Physical Sciences, National Technical University of Athens, Athens, Greece} 
\author{Constantinos Siettos\thanks{Corresponding author: constantinos.siettos@unina.it}} 
\affil{Department of Mathematics and Applications ``Renato Caccioppoli",
Università degli Studi di Napoli Federico II, Naples, Italy}

\date{\today}

\maketitle

\begin{abstract}
We perform one and two-parameter numerical bifurcation analysis of a mechanotransduction model approximating the dynamics of mesenchymal stem cell differentiation into neurons, adipocytes, myocytes and osteoblasts. For our analysis, we use as bifurcation parameters the stiffness of the extracellular matrix and parameters linked with the positive feedback mechanisms that up-regulate the production of the YAP/TAZ transcriptional regulators (TRs) and the cell adhesion area. Our analysis reveals a rich nonlinear behaviour of the cell differentiation including regimes of hysteresis and multistability, stable oscillations of the effective adhesion area, the YAP/TAZ TRs and the PPAR$\gamma$ receptors associated with the adipogenic fate, as well as homoclinic bifurcations that interrupt relatively high-amplitude oscillations abruptly. The two-parameter bifurcation analysis of the Andronov-Hopf points that give birth to the oscillating patterns predicts their existence for soft extracellular substrates ($<1kPa$), a regime that favours the neurogenic and the adipogenic cell fate. Furthermore, in these regimes, the analysis reveals the presence of homoclinic bifurcations that result in the sudden loss of the stable oscillations of the cell-substrate adhesion towards weaker adhesion and high expression levels of the gene encoding Tubulin beta-3 chain, thus favouring the phase transition from the adipogenic to the neurogenic fate. 
\end{abstract}

\section{Introduction}

Mechanotransduction is the process by which cells translate mechanical stimuli into
biochemical signals controlling multiple aspects of cell behaviour including differentiation
\cite{wang2017review}. Mesenchymal stem cells (MSCs) (also known as multipotent marrow stromal
cells) especially, can differentiate into various cells including neurons, adipocytes,
osteoblasts, endothelial cells, pancreatic cells, muscle cells and others according to the
mechanical properties of the tissue on which they are cultured \cite{caplan2009mscs, pittenger1999multilineage,dominici2006minimal}. MSCs clinical significance is undeniable as their differentiation and therapeutic potential has been extensively used in tissue engineering by treating bone traumas, injuries, end-stage organ failure as well as in a number of human diseases such as Parkinson’s disease. Their potential has become all the more valuable in regenerative medicine especially due to the limited number of donations for bone transplantation \cite{ben2010tissue,yoshimi2009self}. On the other hand, undesirable changes in cell’s fate may lead to various diseases as well as abnormal embryonic development. Thus, a comprehensive understanding in the mechanisms that drive cells and consequently multi-cell (tissues) to alter their structure and properties is imperative \cite{pruitt2014mechano}. 

In general, a mechanical stimulus can occur from stretching of the cell, from the flow of a fluid close to their vicinity or the stiffness of the matrix on which they are cultured. Cells attached to relatively soft extracellular matrices (ECM) ($<1KPa$) that resemble brain tissue, experience neurogenic differentiation, stiffer matrices that resemble muscle tissue cause myogenic differentiation, and comparatively rigid matrices ($>10KPa$) that resemble collagenous bone lead to stiff osteogenic cells \cite{engler2006matrix}. Essentially, cells sense the mechanical properties of the substrate and adapt to their surrounding microenvironment by remodelling themselves and differentiate \cite{guilak2009control,trappmann2012extracellular,discher2005tissue}. During differentiation, cells produce one identical cell to the original and another cell that differentiates into a specialized one which is regulated by specific transcription factors \cite{girlovanu2015stem}. Integrins are required proteins for such differentiations. These proteins are essentially receptors that are embedded in the membranes of cells and thus allowing communication between the extracellular matrix (ECM) and the inner stem cell body (intracellular cytoskeleton) \cite{alenghat2002mechanotransduction}. 

Cell adhesion is crucial for the mechanical interaction between a cell and its ECM \cite{eyckmans2011hitchhiker,peng2017mathematical}. The number of chemical bonds and interactions in the surface of the cell is analogous to the degree to which the cell is in contact with the substrate. Cell adhesion is of primary importance in tissue generation and is achieved via materials that act as substrates and thus assist the process of biosynthesis. Static in vitro cell adhesion consists of three stages: attachment of the cell to the substrate, flattening of the cell body and formation of focal adhesion (FA) \cite{khalili2015review}. This is the process where the external tensile force that is created when the transmembrane protein edges of the cell are attached to the ECM causes the connection of the proteins with intra-cellular members called actin filaments that were previously inactive. It has been found that above a threshold adhesive area, adhesion strength and FA stabilize and thus there is no change in mechanical and biochemical stimulations \cite{gallant2005cell}. In addition, cell-to-cell interaction can determine and promote cell differentiation as the mechanical forces that are produced alter the cytoskeletal organization and thus cell’s shape (round or spread and elongated) a factor that largely affects cells' fate by inhibiting or promoting cytoskeletal tension and actomyosin contraction \cite{yourek2007cytoskeletal,muller2013modulating}.

The substrate over which cells are cultured (ECM) is a dynamic structural and multipotent network that continuously undergoes remodelling \cite{theocharis2016extracellular}. Generally, it can be characterized by its ability to resist in deformations due to applied forces, thus its stiffness. In unconstrained bodies Young’s modulus (E) given in Pascals ($Pa$) can be thought of as a measure of stiffness and will be used hereafter to describe the mechanical properties of the cellular microenvironment (ECM). In vitro studies, mimic the matrix elasticity via gels that are coated with collagen to provide the desired adhesion in order for cells to define their microenvironment \cite{engler2006matrix}. The effects of the physical attributes of the ECM such as the matrix stiffness to the MSCs fate have been examined for a long period of time experimentally. Moreover, studies have shown that cells cultured for a period of time on stiff substrates, differentiate into osteogenic cells even after being transferred to softer ones \cite{yang2014mechanical}. This showed that a mechanical memory region exists with regard to the osteogenic fate. 

Mathematical modeling of stem cell dynamics has the ability to make predictions of a biological system’s behavior without resorting to costly and time-consuming experiments that occasionally are overwhelmingly difficult or even impossible with the current technological and theoretical advancements to conduct \cite{piotrowska2008mathematical}. Studies have focused mainly in describing population dynamics and differentiation of cancer stem cells (CSCs) \cite{daukste2009using,johnston2007mathematical,komarova2007effect,turner2009characterization,sun2012stochastic,ainseba2011global}, cell migration, cell proliferation, vascularization, cell death and other dynamic processes with the use of phenomenological modeling techniques \cite{stiehl2011characterization,lei2014mathematical}. Early attempts have focused in modeling the proliferation and differentiation process of stem cells via stochastic modeling based on data of contemporary experimental studies \cite{yakovlev1998stochastic,till1964stochastic}. Generally, mathematical modeling of the process of differentiation especially, must consider the various parameters that regulate cell’s fate including the specific transcription factors, the ECM’s stiffness, cytoskeletal organization, application of mechanical forces and the co-existence of different types of cells \cite{burke2012substrate,mousavi2015role,sun2016computational,paim2018relevant}. 

A deterministic two-dimensional model of PDEs to describe MSC’s differentiation into chondrocytes and osteoblasts during fracture healing has been proposed by Bailon-Plaza and van der Meulen \cite{bailon2001mathematical} and successfully predicted with the use of finite differences the rate of osteogenic growth while considering the ECM synthesis and its time degradation. Lemon et al. \cite{lemon2007mathematical} developed an ODE model based on experimental data that predicts the proliferation and differentiation of MSC’s grown inside artificial porous scaffolds. The model can predict analytically the fate of cells under different levels of oxygen concentration which is associated with higher or lower levels of ECM secretion and thus can aid the study of cell’s behavior inside artificial materials. A model in the form of a PDE for the population balance and a system of ODEs for the consumption of the growth factor which simulates differentiation of cells into connective and non-connective tissues has been proposed by Pisu et al. \cite{pisu2007novel,pisu2008simulation}, which is based upon material balances for growth factors coupled with a mass-structured population balance to simulate cell growth, differentiation and proliferation in vivo or during in vitro cultivation. Stops et al. \cite{stops2010prediction} developed a Computational Fluid Dynamics (CFD) model to describe cell differentiation in scaffolds subjected to mechanical shear strain and fluid flow stresses and results demonstrated that specific combinations of strain and fluid velocity magnitude are associated with specific differentiation lineages. A minimalistic model in the form of ODEs to describe cell differentiation of the osteochondro progenitor cell based on interactions between transcriptional regulators (TRs) has been proposed by Schittler et al. \cite{schittler2010cell} and a complete bifurcation analysis of single cell scale revealed that cell’s fate is regulated by several stimuli as well as their combination and timing. 

The properties and dynamics of epigenetic landscapes (Waddington 1957), in a more general framework have been studied extensively in the past in the form of a toy one-gene or two-gene regulatory network (GRN) with self-activating and mutually-inhibiting transcription factors (TFs) \cite{waddington1957strategy,huang2007bifurcation,ferrell2012bistability,kaity2018reprogramming,wang2011quantifying}. In particular, the process of cell differentiation is modelled with a system of ODEs governing the time evolution of the concentrations of the relevant TFs, while the feedback terms are modelled as Hill functions \cite{ferrell2012bistability}, which account for the multi-stability that is observed in such biological processes \cite{kaity2018reprogramming,wang2011quantifying,suzuki2011oscillatory}. Kaity et al. \cite{kaity2018reprogramming}, studied a two-gene TF regulatory network and showed that one can start from the differentiated state and follow the reverse path to reach the undifferentiated one while observing sustained oscillatory states in the two TFs. In addition, the model provides a theoretical explanation for the phenomenon of transdifferentiation, where one differentiated cell can switch to a different one without passing through the undifferentiated or multipotent state. Suzuki et al. \cite{suzuki2011oscillatory}, proposed a model for a network of cells (that incorporates cell-to-cell interaction), whose protein expressions are regulated by GRNs with five genes and found that stem cells that undergo differentiation consistently exhibit oscillatory patterns. However, studies on the dynamics of actomyosin contractile activity during epithelial morphogenesis have identified that the emergence of sustained oscillations in cell shape is a primarily autonomous mechanism which does not need to rely on cell-to-cell interaction \cite{gorfinkiel2016actomyosin,gorfinkiel2011dynamics}. 

As described above, mathematical models of mechanotransduction have been developed to describe the fate of MSCs driven by external mechanical stimuli, however each model in the literature describes a different experimental observation \cite{till1964stochastic,burke2012substrate,mousavi2015role,sun2016computational}. Recently though a mathematical model \cite{peng2017mathematical} in the form of six ODEs has been proposed predicting the transition fate of MSCs into neurons, adipocytes, myocytes and osteoblasts and their relevant behavior in a continuous range of ECMs stiffness values. The model is able to reproduce the behavior of cells when they are transferred from one substrate to another. Furthermore, it predicts that a mechanical memory region exists for each of the four possible MSC lineages and addresses novel ways with which one can control the relevant fates. In general, differentiation fate can be manipulated by altering three factors: the first substrate stiffness, the duration of the first seeding and the second substrate stiffness. The model predicts that a lower second stiffness value leads to a greater number of MSCs fates through the relevant gene stimulations than with a higher second seeding. In addition, it predicts a number of counter-intuitive dynamic responses. For example, high first stiffness combined with small culture duration and soft second stiffness leads to faster neurogenic stimulation and thus differentiation than a soft substrate alone. However, when the values of first and second stiffness are relatively close to each other the response is the same even if the duration of the first is large. One can also keep both the first and second substrate stiffness values constant and change the duration of the first seeding and see that different possible fates can be produced. Finally, an important finding is that mechanical memory exists provided that the expression level of genes (any of them) are modelled to affect the surface adhesion area in the form of a feedback loop. If this connection is removed from the model, mechanical memory disappears. 

However, apart from the equilibrium states that the model predicts, as we have discussed already above one would have expected to observe sustained oscillations in the concentration of genes. Several experimental studies have also shown the existence of sustained oscillations in the cell adhesion force to the ECM and in the architecture of cells cytoskeleton \cite{gorfinkiel2016actomyosin,gorfinkiel2011dynamics,sanyour2018spontaneous,schillers2010real,zhu2012temporal,hong2014vasoactive,vegh2011spatial}. In this work, we use the arsenal of numerical bifurcation theory to systematically investigate the dynamics of the model in the one and two-parameter space. By doing so we reveal for the first time the emergence of sustained oscillations due to the emergence of Andronov-Hopf bifurcation points and their abrupt disruption due to homoclinic bifurcations in the cell adhesion area and in the levels of gene concentration during differentiation. The oscillations and homoclinic bifurcations are emerging in cells cultured in relatively soft substrates ($\sim0.8kPa$), associated with the adipogenesis of the MSCs but not in stiff substrates where myogenesis and osteogenesis are the dominant differentiation fates. 

\section{Mathematical Model}

The mathematical model has been developed with the use of the following assumptions \cite{peng2017mathematical}. Cells sense the stiffness of the ECM through the adhesion to the substrate. YAP/TAZ are transcriptional regulators which mediate cells perception of the mechanical microenvironment (Yes-associated protein (YAP) and transcriptional coactivator with PDZ-binding domain (TAZ)). It has been found that YAP/TAZ signalling is the key component for tissue morphogenesis. Studies show that stiff ECM correspond to the activation of YAP/TAZ, while soft ECM with its deficiency \cite{dupont2011role,halder2012transduction}. Cells essentially, read the ECM stiffness ($S$) and cytoskeletal tension as levels of YAP/TAZ activity. Additionally, YAP/TAZ is also associated with the expression levels of a number of gene-transcription factors (here TUBB3, PPARG, MYOD1 and RUNX2), which genetically regulate and determine cells’ fate, thus it is also modelled as an upstream factor. The differentiation fates of MSCs are described by the following lineage-specific genes and their corresponding concentrations:

\begin{itemize}
    \item TUBB3 – the gene encoding Tubulin beta-3 chain, associated with a neurogenic fate. It is expressed when cells receive information from a soft substrate ($<1KPa$).
    \item PPARG – peroxisome proliferator-activated receptor gamma, associated with adipogenic fate in soft substrates.
    \item MYOD1 – myogenic differentiation 1, expressed in medium-stiff environments ($\sim10 KPa$).
    \item RUNX2 – runt-related transcription factor 2, leads to osteogenic fate and it is activated in high stiffness environments ($\sim40 KPa$). 
\end{itemize}

\noindent
The following system of coupled ODEs determines the fate of MSCs.

\begin{equation}\label{eqn:eq2.1}
\begin{aligned}
 \frac{d[SAA]}{dt} = k_1 \frac{\displaystyle(\frac{S}{K_1})^{n_1} + (\frac{[TUBB3]}{K_2})^{n_2}}{\displaystyle1+(\frac{S}{K_1})^{n_1} + (\frac{[TUBB3]}{K_2})^{n_2}} + k_2 \frac{\displaystyle(\frac{S}{K_3})^{n_3} + (\frac{[PPARG]}{K_4})^{n_4}}{\displaystyle1+(\frac{S}{K_3})^{n_3} + (\frac{[PPARG]}{K_4})^{n_4}} + \\ k_3 \frac{\displaystyle(\frac{S}{K_5})^{n_5} + (\frac{[MYOD1]}{K_6})^{n_6}}{\displaystyle1+(\frac{S}{K_5})^{n_5} + (\frac{[MYOD1]}{K_6})^{n_6}} + k_4 \frac{\displaystyle(\frac{S}{K_7})^{n_7} + (\frac{[RUNX2]}{K_8})^{n_8}}{\displaystyle1+(\frac{S}{K_7})^{n_7} + (\frac{[RUNX2]}{K_8})^{n_8}} - d_1 [SAA]
  \end{aligned}
 \end{equation}

\begin{equation}\label{eqn:eq2.2}
 \frac{d[YAPTAZ]}{dt} = k_5 [SAA] - d_2 [YAPTAZ]
 \end{equation}

\begin{equation}\label{eqn:eq2.3}
 \frac{d[TUBB3]}{dt} = k_6 \frac{\displaystyle(\frac{[SAA]}{K_9})^{n_9}}{\displaystyle1+(\frac{[SAA]}{K_9})^{n_9} + (\frac{[YAPTAZ]}{K_{10}})^{n_{10}}} - d_3 [TUBB3]
 \end{equation}

\begin{equation}\label{eqn:eq2.4}
 \frac{d[PPARG]}{dt} = k_7 \frac{\displaystyle(\frac{[SAA]}{K_{11}})^{n_{11}}}{\displaystyle1+(\frac{[SAA]}{K_{11}})^{n_{11}} + (\frac{[YAPTAZ]}{K_{12}})^{n_{12}}} - d_4 [PPARG]
 \end{equation}

\begin{equation}\label{eqn:eq2.5}
 \frac{d[MYOD1]}{dt} = k_8 \frac{\displaystyle(\frac{[SAA]}{K_{13}})^{n_{13}}}{\displaystyle1+(\frac{[SAA]}{K_{14}})^{n_{14}} + (\frac{[YAPTAZ]}{K_{13}})^{n_{13}}} - d_5 [MYOD1]
 \end{equation}

\begin{equation}\label{eqn:eq2.6}
 \frac{d[RUNX2]}{dt} = k_9 \frac{\displaystyle(\frac{[SAA]}{K_{15}})^{n_{15}}}{\displaystyle1+(\frac{[SAA]}{K_{16}})^{n_{16}} + (\frac{[YAPTAZ]}{K_{15}})^{n_{15}}} - d_6 [RUNX2]
 \end{equation}
 
 \noindent
 \\
 The nominal values of the model parameters are presented in Table \ref{table:param}.

\begin{table}[ht]
\caption{Nominal values of the model parameters \cite{peng2017mathematical}}
\vspace*{-2mm} 
\centering
\begin{tabular}{c c c c c c} 
\hline
Index & Parameter & Nominal Value & \ \ \ \ \ Index & Parameter & Nominal Value \\ [0.5ex] 
\hline 
1 & $k_1$ & 0.2 & \ \ \ \ \ \ 22 & $n_{13}$ & 20 \\
2 & $k_2$* & 2.2 & \ \ \ \ \ \ 23 & $n_{14}$ & 60\\
3 & $k_3$ & 5 & \ \ \ \ \ \ 24 & $n_{15}$ & 45 \\
4 & $k_4$ & 9 & \ \ \ \ \ \ 25 & $n_{16}$ & 55\\
5 & $k_5$* & 4 & \ \ \ \ \ \ 26 & $K_1$ & 600 \\
6 & $k_6$ & 2.9 & \ \ \ \ \ \ 27 & $K_2$ & 1.1 \\
7 & $k_7$ & 3 & \ \ \ \ \ \ 28 & $K_3$ & 1300\\
8 & $k_8$ & 5 & \ \ \ \ \ \ 29 & $K_4$ &  0.8\\
9 & $k_9$ & 2 & \ \ \ \ \ \ 30 & $K_5$ & 20000\\
10 & $n_1$ & 4 & \ \ \ \ \ \ 31 & $K_6$ & 1\\
11 & $n_2$ & 2 & \ \ \ \ \ \ 32 & $K_7$ & 60000\\
12 & $n_3$ & 6 & \ \ \ \ \ \ 33 & $K_8$ & 1.1\\
13 & $n_4$ & 2 & \ \ \ \ \ \ 34 & $K_9$ & 0.1\\
14 & $n_5$ & 4 & \ \ \ \ \ \ 35 & $K_{10}$ & 0.5\\
15 & $n_6$ & 20 & \ \ \ \ \ \ 36 & $K_{11}$ & 0.89\\
16 & $n_7$ & 6 & \ \ \ \ \ \ 37 & $K_{12}$ & 4\\
17 & $n_8$ & 20 & \ \ \ \ \ \ 38 & $K_{13}$ & 12\\
18 & $n_9$ & 2 & \ \ \ \ \ \ 39 & $K_{14}$ & 3\\
19 & $n_{10}$ & 8 & \ \ \ \ \ \ 40 & $K_{15}$ & 16\\
20 & $n_{11}$ & 2 & \ \ \ \ \ \ 41 & $K_{16}$ & 4.5\\
21 & $n_{12}$ & 8 & \ \ \ \ \ \  & $d_i$ $(i=1,2,..6)$ & 1\\ [1ex] 
\hline 
\multicolumn{2}{c}{*bifurcation parameters}
\end{tabular}
\label{table:param} 
\end{table}

\section{Numerical Bifurcation Analysis: Methods}

We used both MATCONT \cite{dhooge2003matcont}, a numerical bifurcation analysis toolbox  and ``homemade" code  to construct the bifurcation diagrams in the $1D$ parameter space. 
For the computation of steady states we used the Newton-Raphson method augmented by the pseudo-arc-length continuation method \cite{keller1977numerical}. MATCONT uses the Moore-Penrose technique \cite{dhooge2003matcont} for the continuation of solutions past turning points.
As we show and explain due to the form of the nonlinear equations, the Newton-Raphson method fails to detect certain unstable branches of equilibria in the full six ODE model, whose existence was dictated by the outcomes of the bifurcation analysis. 
This is due to the lack of a ``good" initial guess close to the fixed (unstable) point. In these cases, in order to facilitate the analysis, we first performed a model reduction resulting in an one ODE as we describe below. 

As we mentioned, for our computations, we used the pseudo-arc-length continuation method that for the completeness of the presentation we present below together with that for the computation of limit cycles. \\
Consider the continuous time, autonomous nonlinear system

\begin{equation}\label{eqn:eq3.1}
\frac{dx}{dt} = f(x, p),\ \ \ f: {\mathbb{R}}^m \times \mathbb{R}\rightarrow {\mathbb{R}}^m 
\end{equation}

\noindent
where $x \in \mathbb{R}$ denotes the state vector, accessible through measurement, is the bifurcation $x \in \mathbb{R^m}$ parameter and $f$ is a smooth function. 
\par
The task is to trace the steady-state bifurcation diagram of the system, which potentially involves open-loop stable and open-loop unstable steady states, and possibly turning points. In numerical bifurcation calculations, starting with two already known equilibrium points $(x_0, p_0)$
and $(x_1, p_1)$, this is practically accomplished by solving the steady-state equations augmented by the linearized pseudo arc-length condition \cite{keller1977numerical}:

\begin{equation}\label{eqn:eq3.2}
g(x, p) = \alpha (x - x_1) + \beta (p - p_1) - \Delta S = 0
\end{equation}

\noindent
where

\begin{equation}\label{eqn:eq3.3}
\alpha \equiv \frac{(x_1 - x_0)^T}{\Delta S}, \ \ \beta \equiv \frac{p_1 - p_0}{\Delta S}
\end{equation}

\noindent
and $\Delta S$ is the pseudo arc-length continuation step. Eq.\ref{eqn:eq3.2} constrains the next steady-state solution $(x^*, p^*)$ of Eq.\ref{eqn:eq3.1} to lie on a hyperplane perpendicular to the tangent of the bifurcation diagram at $(x_1, p_1)$ --approximated through $(\alpha, \beta)$-- at a distance $\Delta S$ from it.

The continuation of solutions can be obtained using an iterative procedure like the Newton-Raphson technique \cite{kelley1995iterative}. Thus, the procedure involves the iterative solution of the following linearized system:

\begin{equation}\label{eqn:eq3.4}
\begin{bmatrix}
\displaystyle
    \nabla f  &  \frac{\displaystyle{df}}{\displaystyle{dp}}  \\
    \alpha  &  \beta 
\end{bmatrix}
\begin{bmatrix}
    dx   \\
    dp 
\end{bmatrix}
= -
\begin{bmatrix}
    f(x, p)   \\
    g(x, p)
\end{bmatrix}
\end{equation}

It can be shown that the Jacobian of the above augmented system is non-singular on regular turning points \cite{keller1977numerical} where $\det(\nabla f) = 0$ and as a consequence Newton-Raphson can go through turning points. The continuation procedure starts with two known steady states say, $(x_0, p_0)$ and $(x_1, p_1)$ that can be detected far from singular points either by time integration of the system until it reaches the steady state or by the implementation of the Newton-Raphson on function $f$. Of course, only stable steady states can be reached with time integration.

If a periodic oscillatory behavior exists then one seeks for solutions which satisfy the following equation:

\begin{equation}\label{eqn:eq3.5}
x(t) = x(t + T)
\end{equation}

\noindent
with $T$ denoting the period of oscillation

Thus, periodic solutions can be computed by solving the boundary value problem

\begin{equation}\label{eqn:eq3.6}
\begin{pmatrix}
    \dot{x} \\
    \dot{T}
\end{pmatrix}
=
\begin{pmatrix}
    f(x,p) \\
    0
\end{pmatrix}
, \ \ \ \ \
\begin{pmatrix}
    x(0) - x(T) \\
    \emph{h}(x(0), p)
\end{pmatrix}
=0
\end{equation}

\par
\noindent
where $\emph{h}(x(0), p) = 0$ is the so-called phase condition (also called a pinning condition) which factors out the infinite members of the family of periodic solutions. The phase condition enables the computation of the unknown period $T$ by factoring out the translational time-invariance of the problem. For example, such a condition could be the relation $\emph{h}(x(0), p) \equiv x_i(0) - c = 0$ which “pins” , the $\emph{i-th}$ element of the state vector $x$ at $t = 0$ to a prescribed value $c$ or $\emph{h}(x(0), p) \equiv \frac{\displaystyle dx_i(0)}{dt} = 0$  which “pins” $x_i$ at $t = 0$ to be a critical (minimum or maximum) point of $x_i$ or an integral-like condition that takes into account the entire profile of the solution.

\par
The above boundary value problem can be solved using shooting or discretization methods such as finite differences.

\par
For practical reasons one can use a transformation of $t' = \frac{\displaystyle t}{T}$ so that the above problem can be re-written as: 

\begin{equation}\label{eqn:eq3.7}
\begin{pmatrix}
    \dot{x} \\
    \dot{T}
\end{pmatrix}
=
\begin{pmatrix}
    T f(x,p) \\
    0
\end{pmatrix}
, \ \ \ \ \
\begin{pmatrix}
    x(0) - x(1) \\
    \emph{h}(x(0), p)
\end{pmatrix}
=0
\end{equation}

The tracing of the branches of periodic solutions can be achieved using again the linearized pseudo arc-length continuation \cite{keller1977numerical}. Let $(x_0, p_0, T_0)$ and $(x_1, p_1, T_1)$ represent two already known periodic solutions. Then the pseudo-arc length condition reads

\begin{equation}\label{eqn:eq3.8}
N(x, p, T) = \alpha (x - x_1) + \beta (p - p_1) + \gamma(T - T_1) - \delta S = 0
\end{equation}

\noindent
where

\begin{equation}\label{eqn:eq3.9}
\alpha \equiv \frac{(x_1 - x_0)^T}{\delta S}, \ \ \beta \equiv \frac{p_1 - p_0}{\delta S}, \ \ \gamma \equiv \frac{T_1 - T_0}{\delta S}
\end{equation}

\noindent
The continuation of the periodic solutions in once again obtained using an iterative procedure like the Newton-Raphson coupled with pseudo-arc-length.

For our steady state calculations the absolute and relative error for the Newton- Raphson iterations were set equal to $10^{-6}$. 

\begin{figure*}[h!]
\begin{center}
\includegraphics[width=0.35\textwidth]{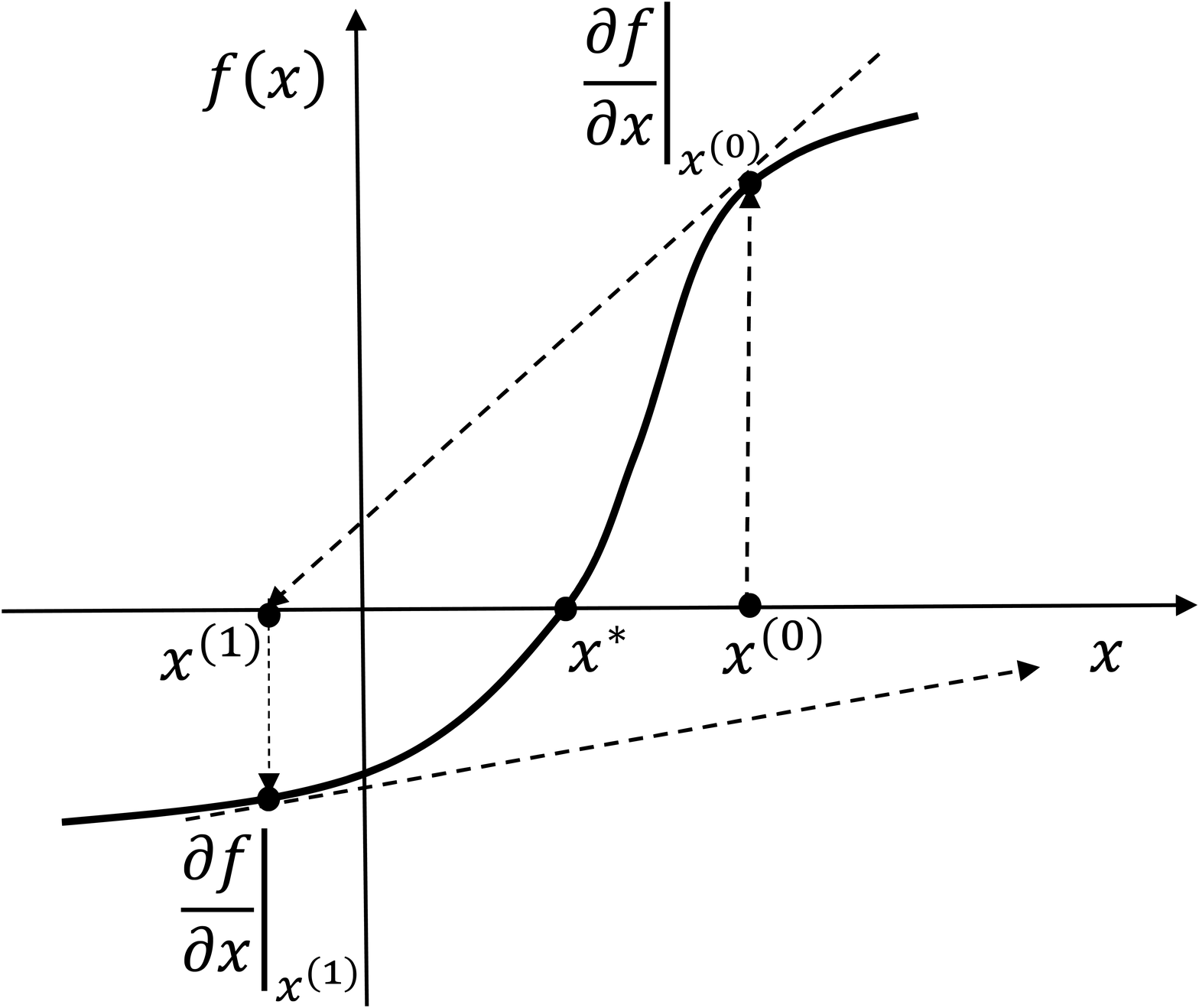} 
\caption{A schematic of a case where Newton-Raphson may fail to converge at the fixed point.}
\label{fig:f1}
\end{center}
\end{figure*}

We used MATCONT as we were aiming at performing a two-parameter bifurcation analysis for the Andronov-Hopf bifurcations. The computation of limit cycles with MATCONT was performed using $20$ mesh points and $4$ collocation points. \\
Regarding the computation of equilibiria, we should note that for the Newton-Raphson to converge on the initial equilibria that are needed for the continuation of solutions, a good initial guess is required even when the Jacobian is non-singular, i.e even when we are far from critical points. One such paradigm in one dimension is illustrated in Figure \ref{fig:f1}.

Choosing $x^{(0)}$ as initial guess for the fixed point $x^{*}$, the first iteration of the Newton-Raphson uses the gradient $\frac{\partial f}{\partial x}\big|_{x^{(0)}}$ to find the next approximation ($x^{(1)}$). Then the gradient $\frac{\partial f}{\partial x}\big|_{x^{(1)}}$ drives the next approximation far away from the desired fixed point $x^{*}$, i.e. the Newton-Raphson fails to detect it. The above one-dimensional paradigm can be extended to the more general case where $f: {\mathbb{R}}^n \rightarrow {\mathbb{R}}$ taking as input a vector $x \in$ ${\mathbb{R}}^n$, where the Hessian

\begin{equation}\label{eqn:eq3.10}
H_{i,j} = \frac{\partial ^2 f}{\partial x_i \partial x_j}
\end{equation}

\noindent
around the fixed point is singular. In the above, $f$ represents one (or more) of the six nonlinear functions of the system of equations to be solved.

\par
Thus, for isolated branches of solutions, i.e. branches of solutions that cannot be traced by continuation, the choice of the initial guess is crucial when one encounters situations as the one illustrated in Figure \ref{fig:f1}. In such cases, one can discretize the space of values of the initial guesses, choosing for example a small step such as $10^{-4}$ or smaller and run the Newton-Raphson algorithm for each one of the initial guesses to converge to all branches of solutions in the desired (physical) range of values. While this can be done for one- or two-dimensional systems, the ``curse of dimensionality” renders computationally prohibitive this task when dealing even with five or six equations. For example, if one has six equations as in our case and seeks to perform an extensive search of the initial guess space, the discretization of the space in $10^4$ values for each one of the variables, will require the implementation of the Newton-Raphson algorithm $10^{24}$ times.

\noindent
Here, to deal with this problem, we constructed a reduced one-dimensional model to search for all steady state solutions in the parametric space. By setting the left-hand-sides of Eq.(\ref{eqn:eq2.3})-(\ref{eqn:eq2.6}) equal to zero we get the following algebraic relations:

\begin{equation}\label{eqn:eq3.11}
[TUBB3] = \frac{k_6}{d_3} \frac{\displaystyle(\frac{[SAA]}{K_9})^{n_9}}{\displaystyle1+(\frac{[SAA]}{K_9})^{n_9} + (\frac{[YAPTAZ]}{K_{10}})^{n_{10}}} 
\end{equation}

\begin{equation}\label{eqn:eq3.12}
[PPARG] = \frac{k_7}{d_4} \frac{\displaystyle(\frac{[SAA]}{K_{11}})^{n_{11}}}{\displaystyle1+(\frac{[SAA]}{K_{11}})^{n_{11}} + (\frac{[YAPTAZ]}{K_{12}})^{n_{12}}} 
\end{equation}

\begin{equation}\label{eqn:eq3.13}
[MYOD1] = \frac{k_8}{d_5} \frac{\displaystyle(\frac{[YAPTAZ]}{K_{13}})^{n_{13}}}{\displaystyle1+(\frac{[SAA]}{K_{14}})^{n_{14}} + (\frac{[YAPTAZ]}{K_{13}})^{n_{13}}} 
\end{equation}

\begin{equation}\label{eqn:eq3.14}
[RUNX2] = \frac{k_9}{d_6} \frac{\displaystyle(\frac{[YAPTAZ]}{K_{15}})^{n_{15}}}{\displaystyle1+(\frac{[SAA]}{K_{16}})^{n_{16}} + (\frac{[YAPTAZ]}{K_{15}})^{n_{15}}} 
\end{equation}

Finally substituting Eq.(\ref{eqn:eq3.11})-(\ref{eqn:eq3.14}) into Eq.(\ref{eqn:eq2.1}) with $\frac{\displaystyle d[SAA]}{\displaystyle dt} = 0$ we get the following equation:

\begin{equation}\label{eqn:eq3.15}
\begin{aligned}
-[SAA] d_1 + k_1 + k_2 + k_3 + k_4 - \frac{\displaystyle k_1}{\displaystyle 1 +(\frac{([SAA]/K_9)^{n_9} k_6}{d_3 K_2 (1+([SAA]/K_9)^{n_9}+([SAA] k_5 / d_2 K_{10})^{n_{10}}})^{n_2}+(S/K_1)^{n_1}} \\ - \frac{\displaystyle k_2}{\displaystyle 1 +(\frac{([SAA]/K_{11})^{n_{11}} k_7}{d_4 K_4 (1+([SAA]/K_{11})^{n_{11}}+([SAA] k_5 / d_2 K_{12})^{n_{12}}})^{n_4}+(S/K_3)^{n_3}} \\ - \frac{\displaystyle k_3}{\displaystyle 1 +(\frac{([SAA] k_5/d_2 K_{13})^{n_{13}} k_8}{d_5 K_6 (1+([SAA]/K_{14})^{n_{14}}+([SAA] k_5 / d_2 K_{13})^{n_{13}}})^{n_6}+(S/K_5)^{n_5}} \\ - \frac{\displaystyle k_4}{\displaystyle 1 +(\frac{([SAA] k_5/d_2 K_{15})^{n_{15}} k_9}{d_6 K_8 (1+([SAA]/K_{16})^{n_{16}}+([SAA] k_5 / d_2 K_{15})^{n_{15}}})^{n_8}+(S/K_7)^{n_7}} = 0
\end{aligned}
\end{equation}

\noindent
\\
This is a function solely of the adhesion area SAA, that can be generally used for the detection of branches of equilibria in the manner described above. However, we should note that the above one-dimensional reduced model can be only used to converge on and trace branches of equilibria but not branches of oscillating solutions. For these, the analysis of the full model is needed.

\section{Numerical Analysis Results}

\subsection{One-parameter Bifurcation Analysis}

\begin{figure*}[ht!]
\begin{center}
\includegraphics[width=0.7\textwidth]{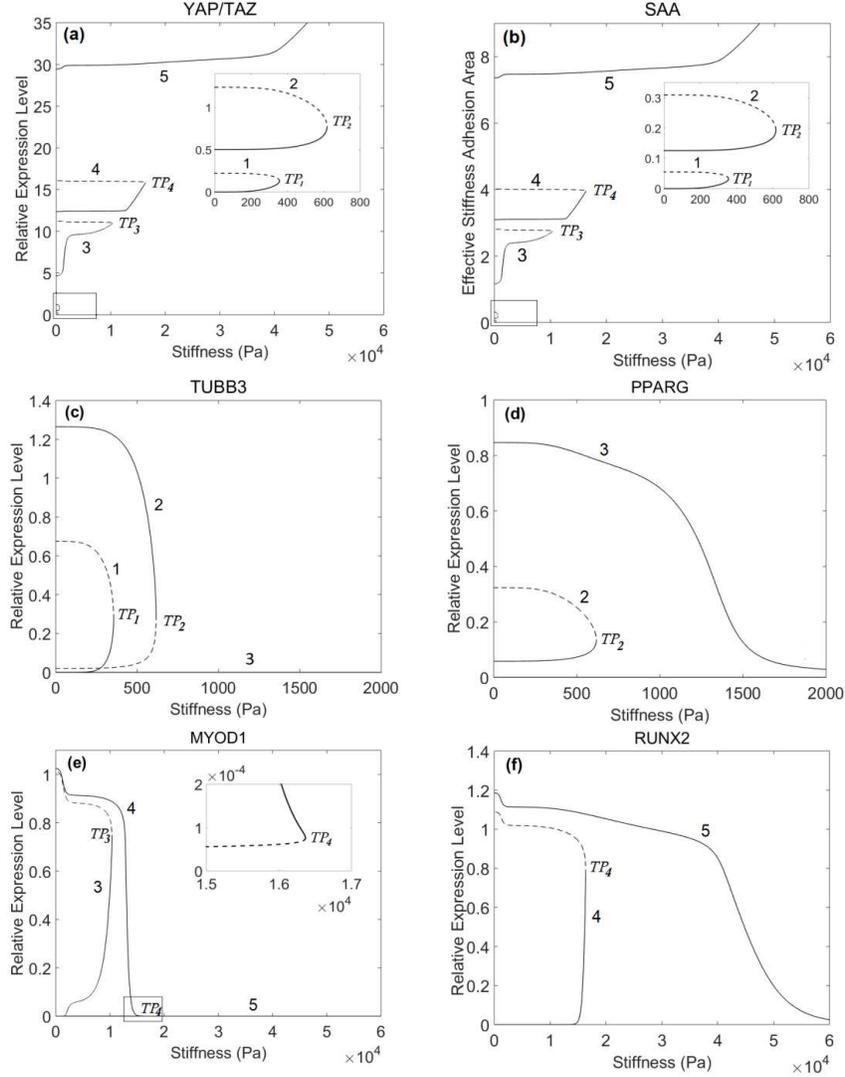} 
\caption{One-parameter bifurcation diagrams of the MSC differentiation model. \textbf{(a)} Relative expression level of YAP/TAZ. \textbf{(b)} Effective stiffness adhesion area. \textbf{(c}-\textbf{f)} Relative expression levels of genes TUBB3, PPARG, MYOD1 and RUNX2 respectively. The bifurcation parameter (x-axis) is the stiffness of the substrate. Solid lines correspond to stable equilibria while dashed lines to unstable equilibria. The branches that are not shown in \textbf{(c}-\textbf{f)} correspond to a value of YAP/TAZ within a range from $10^{-7}$ to $10^{-3}$ (e.g. inset in Figure 1\textbf{(e)}).}
\label{fig:f2}
\end{center}
\end{figure*}

Initially we constructed the one-parameter bifurcation diagrams with respect to the stiffness of the substrate (represented by the parameter $S$) using the same parameter values reported in \cite{peng2017mathematical}. Figures \ref{fig:f2}(a-f) depict the computed bifurcation diagrams constructed using the one-dimensional model. The concentration of the YAP/TAZ regulator with respect to the substrate stiffness is given in Figure \ref{fig:f2}(a) and that of the effective stiffness adhesion area in Figure \ref{fig:f2}(b). Actually here, we have reproduced the bifurcation diagrams presented in \cite{peng2017mathematical} when using the full sets of equations. Thus, the results of the analysis of both the full model with the six equations and the one-dimensional model coincide. There are just five distinct branches of equilibria and no branches of oscillating solutions. The first branch (1) in Figure \ref{fig:f2} corresponds to relatively small values of YAP/TAZ expression and a turning point $(\emph{TP}_1)$  appears at $(S^*_{\emph{TP}_1} = 0.36,\ [YAPTAZ]^*_{\emph{TP}_1} = 0.13,\ [SAA]^*_{\emph{TP}_1} = 0.03,\ [TUBB3]^*_{\emph{TP}_1} = 0.27,\ [PPARG]^*_{\emph{TP}_1} = [MYOD1]^*_{\emph{TP}_1} = [RUNX2]^*_{\emph{TP}_1} \approx 0)$ marking the onset of a solution branch of stationary saddle points. A second saddle-node branch (2) of Figure \ref{fig:f2} appears with a turning point $(\emph{TP}_2)$  at $(S^*_{\emph{TP}_2} = 0.62,\ [YAPTAZ]^*_{\emph{TP}_2} = 0.77,\ [SAA]^*_{\emph{TP}_2} = 0.19,\ [TUBB3]^*_{\emph{TP}_2} = 0.28,\ [PPARG]^*_{\emph{TP}_2} = 0.14,\ [MYOD1]^*_{\emph{TP}_2} = [RUNX2]^*_{\emph{TP}_2} \approx 0)$. The third branch that corresponds to higher levels of YAP/TAZ activity and stable equilibria, reaches a turning point $(\emph{TP}_3)$  at $(S^*_{\emph{TP}_3} = 10.39,\ [YAPTAZ]^*_{\emph{TP}_3} = 10.99,\ [SAA]^*_{\emph{TP}_3} = 2.75,\ [TUBB3]^*_{\emph{TP}_3} \approx 0,\ [PPARG]^*_{\emph{TP}_3} = 8.78 10^{-3},\ [MYOD1]^*_{\emph{TP}_3} = 0.73,\ [RUNX2]^*_{\emph{TP}_3} \approx 0)$ where a path of saddle points emerges with almost constant values of YAP/TAZ. There is also another branch of higher values of concentration, branch (4) with a turning point $(\emph{TP}_4)$  at $(S^*_{\emph{TP}_4} = 16.37,\ [YAPTAZ]^*_{\emph{TP}_4} = 15.83,\ [SAA]^*_{\emph{TP}_4} = 3.96,\ [TUBB3]^*_{\emph{TP}_4} = [PPARG]^*_{\emph{TP}_4} = [MYOD1]^*_{\emph{TP}_4} \approx 0, [RUNX2]^*_{\emph{TP}_4} = 0.77)$. A final fifth curve of stable equilibria appears (branch 5 in Figure \ref{fig:f2}) that corresponds to even higher levels of YAP/TAZ expression. The behavior of the effective adhesion area SAA is analogous to the YAP/TAZ, as at steady state $[SAA]=d_2 [YAPTAZ] / k_5$ (see Figure \ref{fig:f2}(b)). 

Figure \ref{fig:f2}(c) shows the corresponding bifurcation diagram of the relative gene expression of TUBB3. Starting from low (near to zero) levels of the TUBB3 expression when the stiffness remains below $0.36 kPa$ TUBB3 rests in branch (1). As the stiffness increases beyond the first turning point $(S^*_{\emph{TP}_1},\ [TUBB3]^*_{\emph{TP}_1} = 0.27)$ the solution jumps to branch (2). By further increasing the stiffness past the second turning point $(S^*_{\emph{TP}_2},\ [TUBB3]^*_{\emph{TP}_1} = 0.28)$ the third stable branch (3) leads to almost zero concentrations of the gene (i.e. branch (3) is compressed close to zero within a range of $10^{-7}$ to $10^{-4}$). The same behaviour is observed for higher values of the stiffness where again branches (4) and (5) in Figure 2 are compressed to zero values. Figure \ref{fig:f2}(d) shows the corresponding bifurcation diagram for the expression levels of the PPARG gene. Branches (1), (4) and (5) in Figure \ref{fig:f2} are not depicted as they are confined near to zero concentrations. Past the turning point of branch (2) $(S^*_{\emph{TP}_2},\ [PPARG]^*_{\emph{TP}_2} = 0.14)$, the expression of PPARG gene jumps to branch (3) until it reaches a turning point at $(S^*_{\emph{TP}_3},\ [PPARG]^*_{\emph{TP}_3} = 8.78 10^{-3})$ where the curve turns and forms an unstable path of equilibria with values of the concentration close to zero. The bifurcation diagram of the concentration of MYOD1 gene is presented in Figure \ref{fig:f2}(e). Branches (1) and (2) correspond to negligible values of MYOD1 relative expression and thus are not depicted. The value of MYOD1 concentration increases via branch (3) and reaches a turning point $(S^*_{\emph{TP}_4},\ [MYOD1]^*_{\emph{TP}_4} = 87.62 10^{-5})$ of branch (4) is also depicted in the inset of Figure 2(e)). Subsequently, for even higher values of stiffness branch (5) corresponds to approximately zero values of the MYOD1 concentration.

Finally, Figure \ref{fig:f2}(f) illustrates the relative expressions levels of RUNX2 gene that is associated with an osteogenic fate of MSCs. Branches (1) and (2) correspond to negligible values of the gene. However, for higher values of the stiffness ($\sim 15 kPa$), and following the path of branch (4) a turning point emerges at $(S^*_{\emph{TP}_4},\ [RUNX2]^*_{\emph{TP}_4} = 0.77)$. Subsequently, past this turning point the solution jumps to the stable branch (5) characterized by relatively high levels of RUNX2 expression where osteogenic differentiation becomes the dominant fate of the MSCs. 

The bifurcation diagrams of the relative expression levels of the four genes (TUBB3, PPARG, MYOD1, and RUNX2) can explain the relevant “mechanical memory” regions [15]. Starting from zero and by increasing the value of the substrate stiffness, each gene concentration due to the emergence of successive turning points jumps from lower to higher expressions and then abruptly decreases to zero in the case of TUBB3, MYOD1 and PPARG; the expression levels of RUNX2 approach more smoothly zero for higher values of the stiffness ($>60 kPa$). Thus, due to the hysteresis arising from the saddle-node bifurcations, stem cells ‘remember’ past gene concentrations only within a specific range of stiffness values. This region is relatively smaller for neurogenic (Figure \ref{fig:f2}(d)) and adipogenic genes (Figure \ref{fig:f2}(d)) and larger for myogenic (Figure \ref{fig:f2}(e)) and osteogenic (Figure \ref{fig:f2}(f)). For example, we see that osteogenic differentiation (corresponding to high expression levels of RUNX2) persists even if we considerably decrease the stiffness (e.g. from $40 kPa$ to $20kPa$). 

\subsubsection{Oscillations and Homoclinic Bifurcations}

\begin{figure*}[ht!]
\begin{center}
\includegraphics[width=0.7\textwidth]{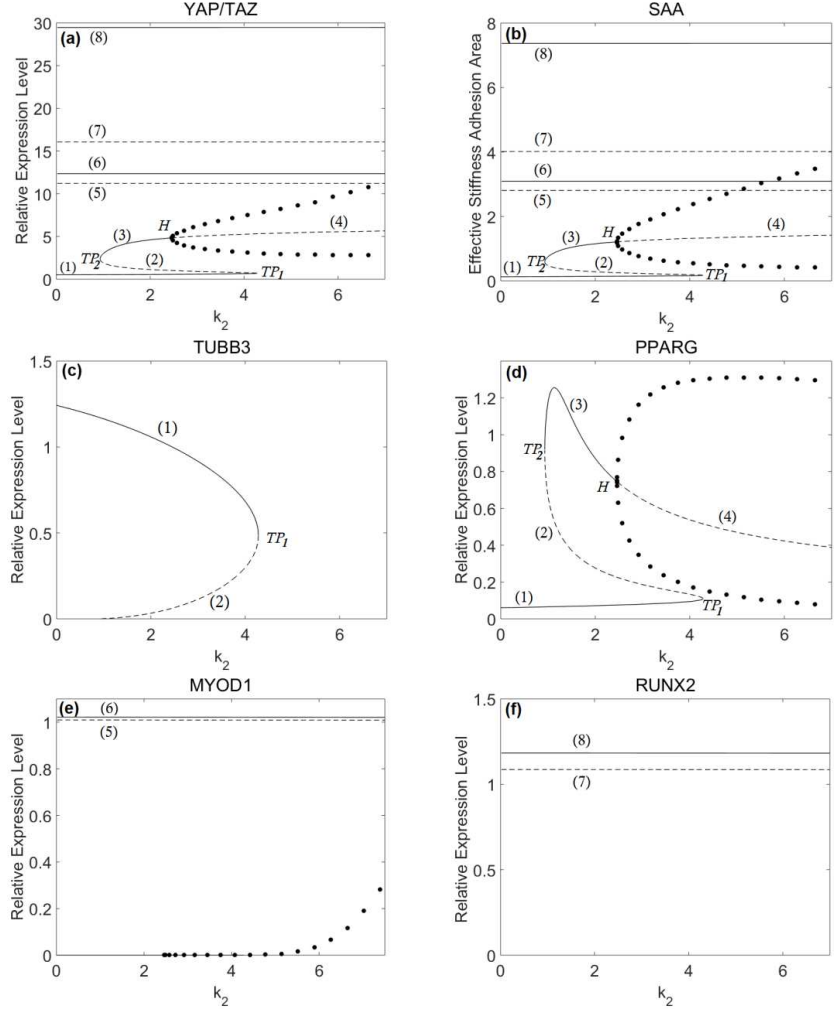}
\caption{One-parameter bifurcation diagrams with respect to parameter $k_2$. \textbf{(a)} Relative expression level of YAP/TAZ. \textbf{(b)} Effective stiffness adhesion area. \textbf{(c-f)} Relative expression levels of genes TUBB3, PPARG, MYOD1 and RUNX2 respectively. The bifurcation parameter (x-axis) is parameter $k_2$. Solid lines correspond to stable equilibria - dashed lines to unstable equilibria and solid dots to maximum and minimum amplitude of oscillation (stable limit cycle)}.
\label{fig:f3}
\end{center}
\end{figure*}

Here, in addition to the study of the dynamical behaviour of the mathematical model with respect to the stiffness of the substrate ($S$), we investigated the system behaviour while changing two other parameters of the model, namely parameters $k_2$ and $k_5$. The parameter $k_2$, which appears in Eq.(\ref{eqn:eq2.1}) is associated with the term that up-regulates the cell adhesion area controlled by the PPAR$\gamma$ receptor agonists. While in the previous analysis the value of this parameter was set $k_2 = 2.2$, here we investigated the dynamic behaviour of the system in the range $[0 - 7.5]$. For the one-parameter bifurcation analysis, we chose to set the stiffness of the substrate at an intermediate value of $S = 500Pa$. At first, we used the full six-dimensional model in order to perform the one-parameter bifurcation analysis. This time, the model exhibits a much more complex behaviour with the emergence of oscillatory patterns (see Figure \ref{fig:f3}). Setting $k_2 = 2.2$ and starting with initial guesses of $([SAA] = 0.14$,\ $[YAPTAZ] = 0.6$,\ $[TUBB3] = 1$,\ $[PPARG] = 0.1$ and $[MYOD1] = [RUNX2] \approx 0)$ we were able to first converge to branch (1) of stable equilibria (see Figures \ref{fig:f3}(a-f)). Then using pseudo-arc length continuation, we traced the branches (2) of unstable equilibria, (3) of stable equilibria and (4) of unstable equilibria. Starting from a stable branch of equilibria (1) with small values of the YAP/TAZ and gradually increasing the value of the parameter we detect the first turning point $(\emph{TP}_1)$ at $(k^*_{2({\emph{TP}_1})} = 4.28,\ [YAPTAZ]^*_{\emph{TP}_1} = 0.696,\ [SAA]^*_{\emph{TP}_1} = 0.17,\ [TUBB3]^*_{\emph{TP}_1} = 0.49,\ [PPARG]^*_{\emph{TP}_1} = 0.11,\ [MYOD1]^*_{\emph{TP}_1} = [RUNX2]^*_{\emph{TP}_1} \approx 0)$. Following the unstable branch (2) in Figure 3, we detected a second turning point $(\emph{TP}_2)$ at $(k^*_{2({\emph{TP}_2})} = 0.933,\ [YAPTAZ]^*_{\emph{TP}_2} = 2.38,\ [SAA]^*_{\emph{TP}_2} = 0.59,\ [TUBB3]^*_{\emph{TP}_2} \approx 0,\ [PPARG]^*_{\emph{TP}_2} = 0.92,\ [MYOD1]^*_{\emph{TP}_2} = [RUNX2]^*_{\emph{TP}_2} \approx 0)$. By further increasing the value of the parameter, the stable branch of equilibria becomes unstable as a supercritical Andronov-Hopf bifurcation $(\emph{H})$ occurs at $(k^*_{2({\emph{H}})} = 2.46,\ [YAPTAZ]^*_{\emph{H}} = 4.84,\ [SAA]^*_{\emph{H}} = 1.21,\ [TUBB3]^*_{\emph{H}} \approx 0,\ [PPARG]^*_{\emph{H}} = 0.75,\ [MYOD1]^*_{\emph{H}} = [RUNX2]^*_{\emph{H}} \approx 0)$ at which stable oscillations emerge. 

The value of the parameter at the bifurcation point is very close to its initial nominal value ($\sim 11 \%$ difference). For even higher values of the parameter, larger amplitude oscillations emerge that continue until reaching a turning point of limit cycles (that is not depicted in the diagrams as it corresponds to large $k_2$ values ($>7$)).

Using the full model, we tried to find other isolated branches of equilibria by time integration trying different initial conditions. By doing so we were able to converge to branches of stable equilibria (6) and (8) of Figure \ref{fig:f3}. However, bifurcation theory dictates the existence of at least two unstable branches of solutions between the branches of stable equilibria (3)-(6) and (6)-(8) (see Figure \ref{fig:f3}). When using the six-ODEs system (Eq.(\ref{eqn:eq2.1})-(\ref{eqn:eq2.6})), we were unable to detect those unstable branches using MATCONT or our Newton-Raphson scheme without a ``good" initial guess. Thus, as explained in the previous section in order to facilitate the detection of all (unstable) equilibrium branches in the range of interest we switched to the one-dimensional model (Eq.(\ref{eqn:eq3.15})). We sought for solutions of SAA in the range $[0 - 15]$ with a discretization step of $10^{-2}$. By doing so, we revealed the existence of branches (5) and (7) shown in Figure \ref{fig:f3}.

\begin{figure*}[h!]
\begin{center}
\includegraphics[width=0.7\textwidth]{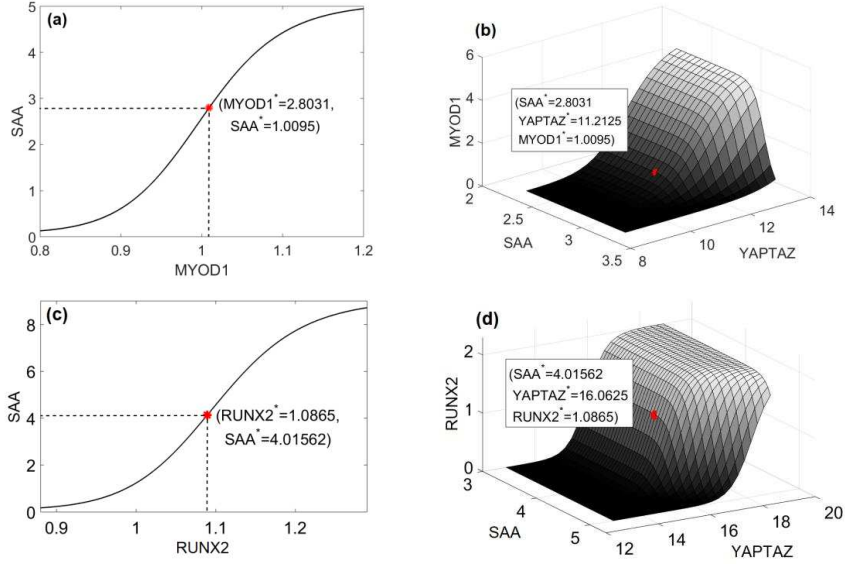}
\caption{Dependence of non-zero valued variables at the unstable branches of equilibria (5) and (7) of Figure \ref{fig:f3} ($S = 0.5 kPa, k_2 = 2.2$). At branch (5) only SAA, YAP/TAZ and MYOD1 differ from zero while at branch (7) only SAA, YAP/TAZ and RUNX2 differ from zero for any practical means. \textbf{(a)} Branch (5): SAA as a function of MYOD1, \textbf{(b)} Branch (5): MYOD1 as a function of SAA, and YAP/TAZ, \textbf{(c)} Branch (7): SAA as a function of RUNX2, \textbf{(d)} Branch (7): RUNX2 as a function of SAA and YAP/TAZ.  The values of the equilibria are also depicted. These sigmoid- like functional reveal why Newton-Raphson fails at detecting these branches when the initial guesses are not sufficient close to the sought equilibria}.
\label{fig:f4}
\end{center}
\end{figure*}

The Newton-Raphson algorithm on the full system could not converge to these branches without a good choice of initial guess due to the reason we explained in the previous section. In particular, when the initial guess is far from the fixed point, the implementation of the Newton-Raphson algorithm on the full system fails to converge to these branches due to the steepness of the sigmoid-like shape of the corresponding functions (see Figure \ref{fig:f4}). For example, regarding the case of branch (5) shown in Figure \ref{fig:f3} (where only three SAA, YAP/TAZ and MYOD1 of the six unknowns-genes differ from zero), the dependence of SAA as a function of MYOD1 (setting TUBB3, PPARG and RUNX2 equal to zero in Eq.(\ref{eqn:eq2.1})) is given by all practical means by:

\begin{equation}\label{eqn:eq4.1}
\begin{aligned}
[SAA] = f_1([MYOD1]) = \frac{k_1}{d_1} \frac{(S/K_1)^{n_1}}{1+(S/K_1)^{n_1}} + \frac{k_2}{d_1}\frac{(S/K_3)^{n_3}}{1+(S/K_3)^{n_3}} \\ + \frac{k_3}{d_1} \frac{(S/K_5)^{n_5} + ([MYOD1]/K_6)^{n_6}}{1+(S/K_5)^{n_5}+([MYOD1]/K_6)^{n_6}} + \frac{k_4}{d_1} \frac{(S/K_7)^{n_7}}{1+(S/K_7)^{n_7}}
\end{aligned}
\end{equation}

\noindent
and the dependence of MYOD1 as a function of SAA, YAP/TAZ (from Eq.(\ref{eqn:eq2.5})) is given by:

\begin{equation}\label{eqn:eq4.2}
\begin{aligned}
[MYOD1] = g_1([SAA],[YAPTAZ]) = \frac{k_8}{d_5} \frac{([YAPTAZ]/K_{13})^{n_{13}}}{1+([SAA]/K_{14})^{n_{14}} + ([YAPTAZ]/K_{13})^{n_{13}}}
\end{aligned}
\end{equation}

\noindent
The plots of $f_1$ and $g_1$ are shown in Figure \ref{fig:f4}(a) and Figure \ref{fig:f4}(b) respectively.

For the unstable branch (7) shown in Figure \ref{fig:f3}, on which again only three of the variables differ from zero (SAA, YAP/TAZ and RUNX2) the dependence of SAA as a function of RUNX2 (setting TUBB3, PPARG and MYOD1 equal to zero in Eq.(\ref{eqn:eq2.1})) is given for all practical means by: 

\begin{equation}\label{eqn:eq4.3}
\begin{aligned}
[SAA] = f_2([RUNX2]) = \frac{k_1}{d_1} \frac{(S/K_1)^{n_1}}{1+(S/K_1)^{n_1}} + \frac{k_2}{d_1}\frac{(S/K_3)^{n_3}}{1+(S/K_3)^{n_3}} \\ + \frac{k_3}{d_1} \frac{(S/K_5)^{n_5}}{1+(S/K_5)^{n_5}} + \frac{k_4}{d_1} \frac{(S/K_7)^{n_7} + ([RUNX2]/K_8)^{n_8}}{1+(S/K_7)^{n_7}+([RUNX2]/K_8)^{n_8}} 
\end{aligned}
\end{equation}

\noindent
and the dependence of RUNX2 as a function of SAA and YAP/TAZ (from Eq.(\ref{eqn:eq2.6})) is given by:

\begin{equation}\label{eqn:eq4.4}
\begin{aligned}
[RUNX2] = g_2([SAA],[YAPTAZ]) = \frac{k_9}{d_5} \frac{([YAPTAZ]/K_{15})^{n_{15}}}{1+([SAA]/K_{16})^{n_{16}} + ([YAPTAZ]/K_{15})^{n_{15}}}
\end{aligned}
\end{equation}

\noindent
The plots of $f_2$ and $g_2$ are shown if Figure \ref{fig:f4}(c) and Figure \ref{fig:f4}(d) respectively.

Thus, when setting $S = 500 kPa$ for a constant value of $k_2$, we can have up to (between the $(\emph{TP}_2)$ and the $(\emph{H})$ point) a total of eight different invariant sets, of which several are stable. Each of them can be derived by appropriate adjustment of the initial conditions of the ODEs such as the initial contact surface between the cell and the ECM. The specific constant values of the YAP/TAZ regulator for branches (5),(6),(7) and (8) are $[YAPTAZ]_{(5)} = 11.21,\ [YAPTAZ]_{(6)} = 12.37,\ [YAPTAZ]_{(7)} = 16.06$ and $[YAPTAZ]_{(5)} = 29.47$ respectively. 

The analysis revealed oscillations in the PPARG gene (Figure \ref{fig:f3}(d)). This clearly illustrates the dominance of this particular gene and thus an adipogenic fate when the cell is cultured to substrate stiffness close to $500 Pa$. Regarding the behaviour of the other genes (Figure \ref{fig:f3}(c,e,f)) the model predicts that they remain in either stable or unstable branches of equilibria and could be expressed by appropriate initial conditions or external stimulus that might cause a sudden change in the stability. 

\begin{figure*}[h!]
\begin{center}
\includegraphics[width=0.7\textwidth]{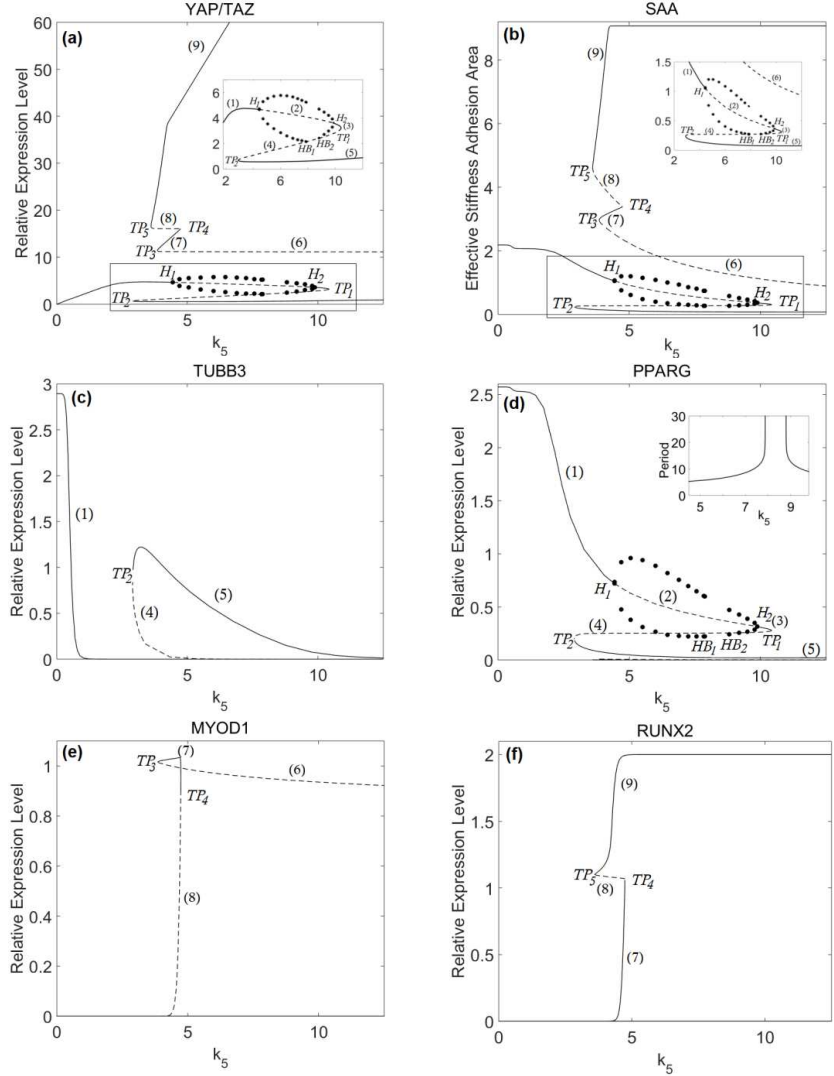}
\caption{One-parameter bifurcation diagrams with respect to the parameter $k_5$. \textbf{(a)} Relative expression level of YAP/TAZ. \textbf{(b)} Effective stiffness adhesion area. \textbf{(c-f)} Relative expression levels of genes TUBB3, PPARG, MYOD1 and RUNX2 respectively. The bifurcation parameter (x-axis) is parameter $k_5$. Solid lines correspond to stable equilibria - dashed lines to unstable equilibria and solid dots to maximum and minimum amplitude of oscillation (stable limit cycle)}.
\label{fig:f5}
\end{center}
\end{figure*}

The existence of oscillations in the genes and adhesion area are also manifested with a perturbation in the parameter $k_5$, associated with the term that switches ``on" the activation of the YAP/TAZ transcriptional factors. The bifurcation diagrams with respect to the parameter $k_5$ are presented in Figure \ref{fig:f5}. Starting from zero and increasing the value of the parameter, the stable branch of equilibria, branch (1), loses its stability through a supercritical Andronov-Hopf bifurcation $(\emph{H}_1)$ at $(k^*_{5({\emph{H}_1})} = 4.44,$ $[YAPTAZ]^*_{\emph{H}_1} = 4.66,$ $[SAA]^*_{\emph{H}_1} = 1.05,$ $[TUBB3]^*_{\emph{H}_1} \approx 0,$ $[PPARG]^*_{\emph{H}_1} = 0.72,$ $[MYOD1]^*_{\emph{H}_1} = [RUNX2]^*_{\emph{H}_1} \approx 0)$, which marks the onset of a branch of stable limit cycles. This branch of stable limit cycles disappears suddenly at a homoclinic bifurcation $(\emph{HB}_1)$ at $(k^*_{5({\emph{HB}_1})} = 7.88,$ $[YAPTAZ]^*_{\emph{HB}_1} = 2.13,$ $[SAA]^*_{\emph{HB}_1} = 0.27,$ $[TUBB3]^*_{\emph{HB}_1} \approx 0,$ $[PPARG]^*_{\emph{HB}_1} = 0.25$ $[MYOD1]^*_{\emph{HB}_1} = [RUNX2]^*_{\emph{HB}_1} \approx 0)$, where the stable limit cycle hits the saddle equilibrium (branch (4) in Figure \ref{fig:f5}). A second Andronov-Hopf supercritical bifurcation point $(\emph{H}_2)$ at $(k^*_{5({\emph{H}_2})} = 9.88,$ $[YAPTAZ]^*_{\emph{H}_2} = 3.61,$ $[SAA]^*_{\emph{H}_2} = 0.36,$ $[TUBB3]^*_{\emph{H}_2} \approx 0,$ $[PPARG]^*_{\emph{H}_2} = 0.31,$ $[MYOD1]^*_{\emph{H}_2} = [RUNX2]^*_{\emph{H}_2} \approx 0)$, where small amplitude oscillations appear until once again another homoclinic bifurcation $(\emph{HB}_2)$ occurs at $(k^*_{5({\emph{HB}_2})} = 8.80,$ $[YAPTAZ]^*_{\emph{HB}_2} = 2.40,$ $[SAA]^*_{\emph{HB}_2} = 0.27,$ $[TUBB3]^*_{\emph{HB}_2} \approx 0,$ $[PPARG]^*_{\emph{HB}_2} = 0.25,$ $[MYOD1]^*_{\emph{HB}_2} = [RUNX2]^*_{\emph{HB}_2} \approx 0)$, where the stable limit cycle hit the saddle equilibria at (branch (4) in Figure \ref{fig:f5}).

By further increasing the value of $k_5$, the system reaches a turning point $(\emph{TP}_1)$ at $(k^*_{5({\emph{TP}_1})} = 10.42,$ $[YAPTAZ]^*_{\emph{TP}_1} = 3.22,$ $[SAA]^*_{\emph{TP}_1} = 0.31,$ $[TUBB3]^*_{\emph{TP}_1} \approx 0,$ $[PPARG]^*_{\emph{TP}_1} = 0.28,$ $[MYOD1]^*_{\emph{TP}_1} = [RUNX2]^*_{\emph{TP}_1} \approx 0)$ and an unstable path appears, branch (4), where the homoclinic bifurcations take place. A second fold point $(\emph{TP}_2)$ is observed at $(k^*_{5({\emph{TP}_2})} = 2.90,$ $[YAPTAZ]^*_{\emph{TP}_2} = 0.69,$ $[SAA]^*_{\emph{TP}_2} = 0.24,$ $[TUBB3]^*_{\emph{TP}_2} = 0.84,$ $[PPARG]^*_{\emph{TP}_2} = 0.199,$ $[MYOD1]^*_{\emph{TP}_2} = [RUNX2]^*_{\emph{TP}_2} \approx 0)$, at which point a branch of stable equilibria continues for even smaller YAP/TAZ values, branch (5) in Figure \ref{fig:f5}. Additionally, in the upper part of the diagram, branch (9) in Figure \ref{fig:f5} shows that the system could abruptly jump from stable state or stable oscillations to very high expression levels. Following this branch, the system reaches a turning point at $(k^*_{5({\emph{TP}_5})} = 3.61,$ $[YAPTAZ]^*_{\emph{TP}_5} = 16.69,$ $[SAA]^*_{\emph{TP}_5} = 4.62,$ $[TUBB3]^*_{\emph{TP}_5} = [PPARG]^*_{\emph{TP}_5} = [MYOD1]^*_{\emph{TP}_5} \approx 0,$ $[RUNX2]^*_{\emph{TP}_5} 1.10)$ and two more at $(k^*_{5({\emph{TP}_4})} = 4.73,$ $[YAPTAZ]^*_{\emph{TP}_4} = 16.02,$ $[SAA]^*_{\emph{TP}_4} = 3.38,$ $[TUBB3]^*_{\emph{TP}_4} = [PPARG]^*_{\emph{TP}_4} \approx 0,$ $[MYOD1]^*_{\emph{TP}_4} = 0.95,$ $[RUNX2]^*_{\emph{TP}_4} = 1.03)$ and $(k^*_{5({\emph{TP}_3})} = 3.86,$ $[YAPTAZ]^*_{\emph{TP}_3} = 11.43,$ $[SAA]^*_{\emph{TP}_3} = 2.96,$ $[TUBB3]^*_{\emph{TP}_3} \approx 0,$ $[PPARG]^*_{\emph{TP}_3} = 0.01,$ $[MYOD1]^*_{\emph{TP}_3} = 1.02,$ $[RUNX2]^*_{\emph{TP}_3} \approx 0)$, consecutively.

Although the relevant bifurcation diagram of the effective substrate adhesion area SAA with respect to the parameter $k_5$ (Figure \ref{fig:f5}(b)) is similar, we notice that the upper branch of stable equilibria (9), reaches a constant value $([SAA] \approx 9.07)$ for $k_5 > 4.31$  whereas in Figure \ref{fig:f5}(a), this branch increases intensely. Interestingly, the system exhibits multiple equilibrium solutions (up to seven) as well as oscillations in an interval of $k_5$-values that are very close to its nominal value $(k_5 = 4)$. Appropriate initial conditions are required in order for the system to exhibit the values of one of these possible states. 

The expression levels of the four genes under investigation are presented in Figures \ref{fig:f5}(c-f). The model predicts that the oscillations of the adhesion area will cause relevant signals solely on the PPARG gene (Figure \ref{fig:f5}(d) - branches (2) and (3)) which is associated with an adipogenic fate. On the contrary, the remaining three gene expression levels will remain in either stable or unstable equilibrium branches without the appearance of oscillations (Figures \ref{fig:f5}(c,e,f)). 

For the sake of completeness, the bifurcation diagrams with respect to $k_5$ present a wide range of parameter values, however $k_5$ might have little difference from its original estimated value $(k_5 = 4)$. Interestingly, the model predicts that the advent of oscillating genes appears with just a small change of the parameter’s nominal value. It is noted that the oscillatory patterns that appeared for a specific range of the parameter can be also computed with a reduced system of ODEs. More specifically, we observe that the oscillations for three out of the four genes (TUBB3, MYOD1 and RUNX2) are suppressed to zero. In such an event, the associated branches (2) and (3) upon which the oscillations occur can be computed numerically by omitting Eq.\ref{eqn:eq2.3},\ref{eqn:eq2.5},\ref{eqn:eq2.6}. 

The homoclinic bifurcations that stops suddenly the oscillatory behavior in Figure \ref{fig:f5}(d), drive the system to the stable branch (5) of Figure \ref{fig:f5}. This transition leads the system to high concentrations of the TUBB3 gene which in turned ‘on’ in soft stiffness environments and causes a neurogenic differentiation. 

\subsection{Two-Parameter Bifurcation Analysis}

To trace the Andronov-Hopf and homoclinic bifurcations that mark the onset of sustained oscillations and their sudden loss, we conducted a two-parameter bifurcation analysis with respect to the stiffness $S$ and the model parameters $k_2$ and $k_5$. 

\begin{figure*}[ht]
\begin{center}
\includegraphics[width=0.85\textwidth]{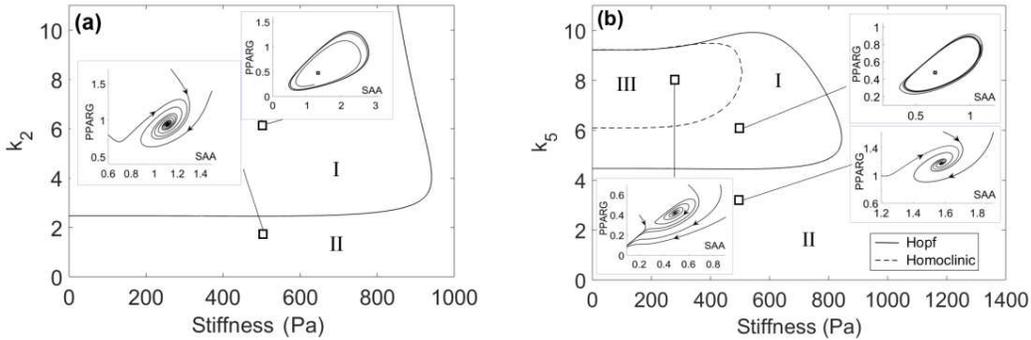}
\caption{Two-parameter bifurcation analysis of the Andronov-Hopf (solid lines) and homoclinic bifurcations (dotted line) \textbf{(a)} with respect to $k_2$ and $S$ (see Figure \ref{fig:f3}) and \textbf{(b)} with respect to $k_5$ and $S$ (see Figure \ref{fig:f5}). Inside area I, the phase portraits for \textbf{(a)} $(S = 500Pa$,\ $k_2 = 6)$ and \textbf{(b)} $(S = 500Pa$,\ $k_5 = 6)$ are presented depicting small-amplitude stable limit cycles with the unstable equilibrium points inside. Inside regime II, the phase portraits \textbf{(a)} $(S = 500Pa$,\ $k_2 = 1.8)$ and \textbf{(b)} $(S = 500Pa$,\ $k_5 = 3)$ are presented depicting the stable spirals. Regime III in \textbf{(b)}, encircled by the dashed-line of homoclinic bifurcations corresponds to the area of unstable equilibria; a relevant phase portrait (unstable spiral) for $(S = 300Pa$,\ $k_5 = 8)$ is also shown.}
\label{fig:f6}
\end{center}
\end{figure*}

The results of this analysis are depicted in Figure \ref{fig:f6}. Figure \ref{fig:f6}(a) shows the two-parameter bifurcation diagram with respect to $k_2$ and $S$ (see Figure \ref{fig:f3}). Figure \ref{fig:f6}(b) shows the two-parameter bifurcation diagram with respect to $k_5$ and $S$ of the $(\emph{H}_1)$  Andronov-Hopf point (see Figure \ref{fig:f5}). Solid lines correspond to Andronov-Hopf points. The area I represent the regime where stable oscillations may be observed, while area II the regime of equilibria. For regime I, characteristic phase portraits that depict the limit cycles and the relevant unstable equilibrium points inside are presented for $(S = 500Pa,\ k_2 = 6)$ (Figure \ref{fig:f6}(a)) and for $(S=500Pa,\ k_5 = 6)$ (Figure \ref{fig:f6}(b)) while for the regime II the phase portraits for the steady-states (stable spirals) are presented for $(S = 500Pa,\ k_2 = 2)$ (Figure \ref{fig:f6}(a)) and $(S = 500Pa,\ k_5 = 3)$ (Figure \ref{fig:f6}(b)). In Figure \ref{fig:f6}(b), the dashed line represents the points of homoclinic bifurcations, while in regime III a phase portrait for the unstable equilibria (unstable spiral) is presented for $(S = 300Pa,\ k_5 = 8)$ (Figure \ref{fig:f6}(b)). We observe that Andronov-Hopf bifurcations, and thus oscillations, occur for ECM’s stiffness values up to $S \approx 900Pa$. These relatively low-to-intermediate values of the stiffness correspond to the oscillation of the PPARG gene concentrations, associated with the adipogenic fate. Our analysis revealed no oscillatory patters regarding a myogenic or osteogenic differentiation.

\section{Discussion}

One of the fundamental challenges in developmental biology regards the understanding and modelling of mechanisms pertaining to the dynamics of the up- and down-regulation of lineage-specific genes, a process driven by cell‐autonomous mechanisms or by exogenous factors such as the interaction with neighbouring cells and/or the extracellular matrix (ECM) \cite{gorfinkiel2016actomyosin}. Since the introduction of the Waddington’s epigenetic landscape of bifurcating valleys in 1957 \cite{waddington1957strategy} describing in a phenomenological way how gene regulation and mutual interaction pertain to cell differentiation and fate \cite{ferrell2012bistability,wang2011quantifying}, there has been an increasingly interest in developing mathematical models that aspire to explain the underlying mechanisms \cite{kaity2018reprogramming}.

Most of the studies have focused on the modelling of gene regulatory networks (GRNs) -usually with a minimum number of them-that govern the level of gene expressions, i.e. emphasizing the role of cell-autonomous mechanisms to the fate of cells \cite{kaity2018reprogramming,wang2011quantifying,suzuki2011oscillatory,gorfinkiel2016actomyosin,gorfinkiel2011dynamics,wang2008potential}. Thus, the investigation of the mechanisms that pertain to the emergence of cell transitions is performed through bifurcation analysis \cite{huang2007bifurcation,ferrell2012bistability,kaity2018reprogramming,rabajante2015branching,verd2014classification,wang2010potential,wang2011quantifying,li2013quantifying}. However, experimental studies also suggest the importance of external mechanical stimuli, and in particular the importance of the stiffness of the extracellular matrix on MSCs fate determination \cite{guilak2009control,trappmann2012extracellular,discher2005tissue}. Mathematical models that incorporate this external stimulus -without however considering the effects of regulatory factors- include a 3D mechanosensing model proposed by Mousavi et al. \cite{mousavi2015role} and a model describing cell differentiation in tissue regeneration developed by Burke et al. \cite{burke2012substrate}.

The regulatory network proposed recently by Peng et al. \cite{peng2017mathematical} examines the fate of cells driven by the stiffness of the ECM and the transcriptional factors YAP and TAZ (responsible for cell’s perception of the mechanical micro-environment) that interact with four other TFs-genes in a continuous range of stiffness values. The authors address a phenomenological model consisting of six ODEs, whose parameters are tuned by experimental data and perform a one-parameter bifurcation analysis with respect to the stiffness of the ECM. By doing so, they identified and quantified novel mechanical memory regions that arise due to fold bifurcations that have a direct impact on cell fate transitions. Re-seeding MSCs into substrates of different stiffness and altering the culture duration leads to a number of possible fates, a finding that can greatly contribute to the understanding and enhancement of regenerative therapeutic treatments. 

Here, we extend the numerical bifurcation analysis of the model proposed by Peng et al. \cite{peng2017mathematical}. First, we provided a model reduction in one dimension that allowed us to identify steady states whose detection with the full six-ODE model, would be, as explained, an overwhelming task. We performed both one- and two-parameter numerical bifurcation analysis with respect to the parameters associated with the positive feedback mechanisms that activate the expression of the YAP/TAZ TRs and the cell adhesion area. The one-parameter bifurcation analysis revealed the emergence of sustained oscillations that arise from Andronov-Hopf bifurcations, in the adhesion area, the YAP/TAZ factors and the PPAR$\gamma$ receptors. We found that soft to medium stiffness substrates associated with neurogenic and adipogenic fates are required for the manifestation of oscillatory signals. The two-parameter bifurcation analysis revealed that the bifurcation points and oscillatory patterns in the adhesion area of cells and in concentration of genes emerge on relatively soft stiffness environments $(< 1 kPa)$. Importantly, we found a new mechanism of cell fate transitions due to the existence of homoclinic bifurcations. These stop abruptly the oscillations and drive the system to high concentrations of genes that favour the neurogenic or the adipogenic fate. Slight changes on the parameters studied here, can drive the system for a stable state to an oscillating one and enable changes in ultimate cell fate. Our findings contribute to the understanding of how oscillations in TFs/genes and effective adhesion area of cells can emerge and terminated. To the best of our knowledge this is the first time that this mechanism due to homoclinic bifurcations is shown for cell fate transitions.

Spontaneous oscillations in cell adhesion have been observed in several experimental studies \cite{gorfinkiel2016actomyosin,gorfinkiel2011dynamics,sanyour2018spontaneous,schillers2010real,zhu2012temporal,hong2014vasoactive,vegh2011spatial}, including studies on the behaviour of vascular smooth cells and cerebral endothelial cells. Our findings can drive the design of new experimental studies in order to reproduce the oscillatory patterns and their abrupt disappearance that we have found in MSCs cultured in soft-to-medium stiffness substrates and particularly express high concentrations of the peroxisome proliferator-activated receptor gamma (PPAR$\gamma$) gene which enables MSCs to differentiate into adipocytes.


\bibliographystyle{unsrtnat}
\bibliography{text_paper}

\end{document}